\documentclass[letterpaper, 10 pt, conference]{ieeeconf}  

\IEEEoverridecommandlockouts                              

\usepackage[utf8]{inputenc}

\usepackage{amsmath} 
\usepackage{amssymb}  
\usepackage{graphicx,verbatim}
\usepackage{mdframed}
\usepackage{arydshln}

\usepackage[ruled,vlined]{algorithm2e}
\usepackage{setspace}

\newtheorem{theorem}{\bfseries Theorem}
\newtheorem{definition}{\bfseries Definition}
\newtheorem{lemma}{\bfseries Lemma}
\newtheorem{assumption}{\bfseries Assumption}

\providecommand{\iprod}[2]{\ensuremath{\left\langle #1,\,#2  \right\rangle}}
\providecommand{\norm}[1]{\ensuremath{\left\lVert#1\right\rVert }}
\providecommand{\mnorm}[1]{\ensuremath{\left\lvert#1\right\rvert}}

\def\R{\mathbb{R}}
\def\H{\mathcal{H}}
\def\B{\mathcal{B}}

\def\G{\mathcal{G}}

\def\F{\mathcal{F}}

\def\X{\mathcal{X}}

\def\V{\mathcal{V}}

\def\E{\mathcal{E}}
\def\S{\mathcal{S}}

\def\a{\textbf{a}}
\def\b{\textbf{b}}

\title{\LARGE \bfseries Byzantine Fault-Tolerance in Decentralized Optimization \\ under Minimal Redundancy}

\author{Nirupam Gupta$^{\dag}$ \hspace{0.5 in } Thinh T. Doan$^{\ddag}$ \hspace{0.5 in } Nitin H. Vaidya$^{\dag}$
\thanks{*This work was supported by the Army Research Laboratory under Cooperative Agreement W911NF-17-2-0196, and by the National Science Foundation award 1842198.}
\thanks{$^{\dag}$Department of Computer Science, Georgetown University, Washington DC. Emails: {\tt\small nirupam115@gmail.com} and {\tt\small nitin.vaiyda@georgetown.edu}.}%
\thanks{$^{\ddag}$Department of Electrical and Computer Engineering, Virginia Tech. Email: {\tt\small thinhdoan@vt.edu}.}%
}

\begin{document}

\maketitle

\begin{abstract}
This paper considers the problem of Byzantine fault-tolerance in multi-agent decentralized optimization. In this problem, each agent has a local cost function. The goal of a decentralized optimization algorithm is to allow the agents to cooperatively compute a common minimum point of their aggregate cost function. We consider the case when a certain number of agents may be Byzantine faulty. Such faulty agents may not follow a prescribed algorithm, and they may share arbitrary or incorrect information with other non-faulty agents. Presence of such Byzantine agents renders a typical decentralized optimization algorithm ineffective. We propose a decentralized optimization algorithm with provable {\em exact fault-tolerance} against a bounded number of Byzantine agents, provided the non-faulty agents have a {\em minimal redundancy}.
     
\end{abstract}

\section{\bfseries Introduction} 
\label{sec:intro}
\begin{figure}[htb!]
\centering
\centering \includegraphics[width=0.35\textwidth, height=0.17\textwidth]{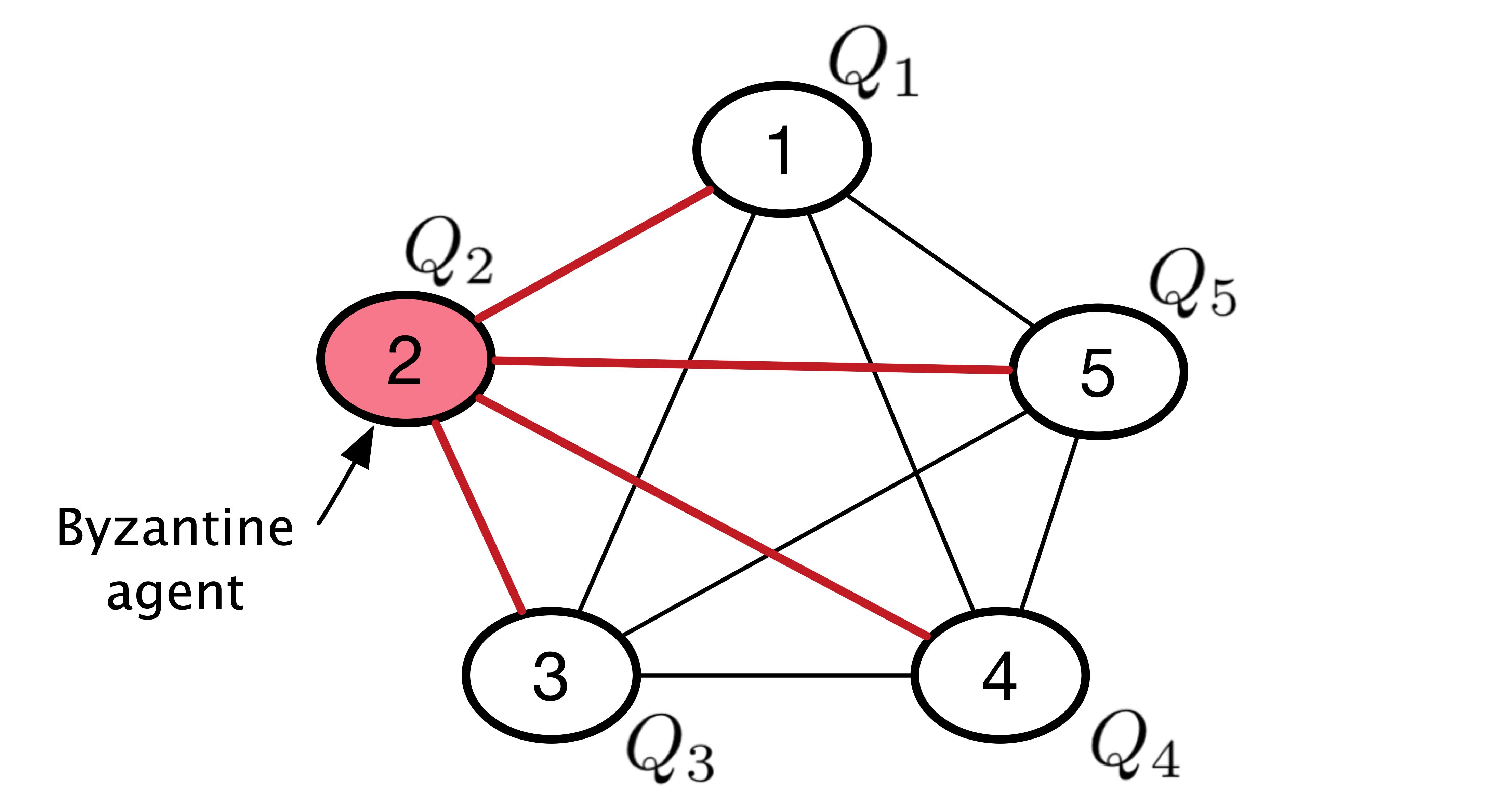} 
\caption{\footnotesize{\it The system architecture.}}
\label{fig:sys}
\end{figure}

In this paper we consider a multi-agent optimization problem defined over a peer-to-peer network of $n$ agents. The network can be modeled by a complete graph $\G = (\V, \, \E)$, as illustrated in Fig. \ref{fig:sys} for $n = 5$, where $\V = \{1,\ldots,n\}$ is the set of agents and $\E = \V \times \V$ is the set of communication links between the agents. In this problem, each agent $i$ has a convex and differentiable cost function $Q_i: \R^d \to \R$. In the fault-free setting, i.e., when all the agents correctly follow a specified algorithm, the goal of the agents is to cooperatively compute a minimum point of the aggregate of their cost functions, i.e., a point
$x^*$ that satisfies
\begin{align}
    x^* \in \arg \min_{x \in \R^d} ~ \sum_{i \in \V } Q_i(x) . \label{eqn:orig_obj}
\end{align}
~

In the past, the above decentralized optimization problem has gained significant attention due to its broad applications~\cite{boyd2011distributed, nedic2009distributed}. Notable applications of decentralized optimization include swarm robotics~\cite{raffard2004distributed}, multi-sensor networks~\cite{rabbat2004distributed}, and distributed machine learning~\cite{boyd2011distributed}.  However, most prior work assumes the fault-free setting where all the agents follow a specified algorithm correctly, e.g., see~\cite{nedic2009distributed} and references therein. We consider a setting where some of the agents may be Byzantine faulty~\cite{lamport1982byzantine}. \\

The problem of decentralized optimization with Byzantine faulty agents was first introduced by Su and Vaidya~\cite{su2016fault}. Byzantine faulty agents may behave arbitrarily, and their identity is a priori unknown to the non-faulty agents~\cite{lamport1982byzantine}. In particular, the Byzantine faulty agents
may collude and share incorrect information with other non-faulty agents in order to corrupt the output of a decentralized optimization algorithm. 
For example, consider an application
of decentralized optimization to the case of sensor networks where there are multiple sensors, and each sensor partially observes a common {\em object} in order to collectively identify the object~\cite{rabbat2004distributed}. However, faulty sensors may share information corresponding to arbitrary incorrect observations to prevent the non-faulty sensors from correctly identifying the object~\cite{su2018finite}. In the case of decentralized learning, faulty agents may share information based upon {\em mislabelled} or arbitrary concocted data points to prevent the non-faulty agents from learning a {\em good} classifier~\cite{charikar2017learning, chen2017distributed, xie2019zeno}.\\

We consider the multi-agent decentralized optimization problem in the presence of up to $f$ Byzantine faulty agents. Our goal is to design a decentralized optimization algorithm that allows all the non-faulty agents to compute a minimum of the aggregate cost of just the non-faulty agents~\cite{gupta2020fault_podc}. In particular, we consider the problem of {\em exact fault-tolerance} defined below. We denote the cardinality of a set $\H$ by  $\mnorm{\H}$. \\

\begin{definition}[Exact fault-tolerance]
\label{def:eft}
\!Let $\H\! \subseteq\! \V$, with $\mnorm{\H} \geq n-f$, be the set of non-faulty agents. A decentralized optimization algorithm is said to have {\em exact fault-tolerance} if it allows each non-faulty agent to compute $x^*_{\H}$ defined as
\begin{align}
    x^*_\H \in \arg \min_x \sum_{i \in \H} Q_i(x). \label{eqn:hon_obj}
\end{align}
\end{definition}
~

Since the identity of the Byzantine faulty agents is a priori unknown to the non-faulty agents, in general, {\em exact fault-tolerance} is unachievable~\cite{su2016fault}. In particular, exact fault-tolerance is impossible unless the non-faulty agents satisfy the {\em $2f$-redundancy} property, proposed in~\cite{gupta2020fault_podc, gupta2020resilience}, defined formally as follows.\\






\begin{definition}[{\em $2f$-redundancy}]
\label{def:2t_red}
A set of non-faulty agents $\H$, with $\mnorm{\H} \geq n-f$, are said to have {\em $2f$-redundancy} if for each subset $\S \subseteq \H$ with $\mnorm{\S} \geq n-2f$, 
the following condition holds true
$$\arg \min_{x \in \R^d} \, \sum_{i \in \S} Q_i(x) = \arg \min_{x \in \R^d} \, \sum_{i \in \H} Q_i(x).$$
\end{definition}
~

The \mbox{$2f$-redundancy} property implies that a minimum of the aggregate cost of any $n-2f$ non-faulty agents is also a minimum of the aggregate cost of all the non-faulty agents, and vice-versa. Although the {$2f$-redundancy} property may appear somewhat technical at this point, we note that, in many practical applications, redundancy in cost functions occurs naturally~\cite{gupta2020fault_podc, gupta2019byzantine}. Indeed,
such redundancy is easily realized in practical applications such as distributed sensing~\cite{gupta2019byzantine, chong2015observability}, and homogeneous distributed learning~\cite{blanchard2017machine, gupta2020byzantine}. \\

We propose a decentralized optimization algorithm that can achieve an exact fault-tolerance, provided the necessary condition of {$2f$-redundancy} is satisfied and the fraction of Byzantine faulty agents is bounded. Similar to a typical fault-free decentralized optimization algorithm~\cite{bertsekas1989parallel, nedic2009distributed}, our proposed algorithm is an iterative method, which is implemented synchronously by the non-faults agents using inter-agent interactions. Specifically, each non-faulty agent maintains a local variable as an estimate of the non-faulty optimal solution $x^*_\H$ defined in~\eqref{eqn:hon_obj}. The non-faulty agents then iteratively update their variables while only exchanging messages with their neighboring agents. In addition, 
in our algorithm the non-faulty agents use a vector filter named {\em Comparative Elimination} (CE) filter, that we propose, to mitigate the detrimental impact of potentially incorrect values shared by the Byzantine faulty agents. 
The proposed CE filter is the key component of our algorithm to achieve an exact fault-tolerance.
A formal description of our algorithm and its fault-tolerance property are presented in Section~\ref{sec:algo}.\\



Next, we present a summary of our contributions, and then discuss the related work.

\subsection{\bfseries Summary of our contributions}
We show that our proposed decentralized optimization algorithm has provable {\em exact fault-tolerance} in a complete peer-to-peer network, if certain assumptions about the non-faulty agents' cost functions are satisfied. 
Our fault-tolerance result can be informally stated as follows. Please refer to Section~\ref{sec:ft} for details. \\

{\bf Theorem} ({\em Informal}): Suppose that the non-faulty agents' costs have {\em $\mu$-Lipschitz} gradients, and the average cost function of the non-faulty agents is {\em $\lambda$-strongly convex}. Our algorithm has {\em exact fault-tolerance} if the non-faulty agents satisfy the necessary condition of $2f$-redundancy, and 
\[f/(n-f) < \left(1 - \sqrt{1 + (\lambda/\mu)} \right)^2.\]


\subsection{\bfseries Related work}
The prior work on fault-tolerance in multi-agent decentralized optimization~\cite{su2016fault, su2020byzantine, sundaram2018distributed} consider {\em approximate} fault-tolerance in which the agents compute an approximate minimum of the non-faulty aggregate cost. Specifically, the decentralized algorithms proposed in these works output a point that minimizes a {\em non-uniformly} {weighted} aggregate of the non-faulty cost functions, instead of the actual uniformly weighted aggregate defined in \eqref{eqn:hon_obj}. Moreover, these works only consider {\em univariate} cost functions, i.e., $x\in\mathbb{R}$. On the other hand, we consider the more general {\em multivariate} cost functions, i.e., $x\in\mathbb{R}^{d}$ for $d\geq 1$, and present results for an {\em exact} fault-tolerance. We note that it is not obvious whether we can apply the work in \cite{su2016fault, su2020byzantine, sundaram2018distributed} for the multivariate cost functions, even in the context of approximate fault-tolerance problems. Indeed, the norm-filter proposed in this paper is new and fundamentally different from the one in the existing literature.  \\ 

There are works on {\em approximate} fault-tolerance for multivariate cost functions~\cite{kuwaranancharoen2020byzantine, su2016robust, yang2017byrdie}. However,~\cite{su2016robust} and~\cite{yang2017byrdie} consider degenerate cases of the multi-agent optimization problem defined in~\eqref{eqn:orig_obj}. In particular,~\cite{su2016robust} assumes that the agents' cost functions are linear combinations of a common set of {\em basis functions}, and~\cite{yang2017byrdie} assumes that the agents' costs can be {\em decomposed} into independent {\em univariate strictly convex} functions. Similar to~\cite{su2016fault}, the algorithms proposed in~\cite{su2016robust, yang2017byrdie} output a minimum of a {\em non-uniformly} weighted aggregate of the non-faulty cost functions. On the other hand, the algorithm in~\cite{kuwaranancharoen2020byzantine} outputs a point in a proximity of a true minimum. Unlike these works, we are interested in {\em exact} fault-tolerance under the necessary condition of $2f$-redundancy. \\

Prior work~\cite{su2018finite} considers the problem of decentralized linear sensing, a special case of decentralized optimization studied in this paper. To guarantee an exact  fault-tolerance, in addition to the $2f$-redundancy their proposed algorithm relies on an assumption on the {observations} of non-faulty agents. When applied to the specific decentralized linear sensing problems, our algorithm achieves exact fault-tolerance under weaker assumptions, that is, we assume the $2f$-redundancy property and the fraction of faulty agents $f/(n-f)$ being smaller than a threshold determined the {\em condition number} 
of the non-faulty {\em observation matrix}.\\

Finally, it is worth pointing out that our fault-tolerance result is only proven for the case of a complete network, where all the agents can interact with each other. However, the problem of Byzantine fault-tolerant decentralized optimization considered in this paper is nontrivial, and it is not obvious how to apply the existing techniques for solving this problem even for the special case of complete network. We, therefore, take a first step by considering this special case. An extension of our work for the more general incomplete network is interesting, which we leave for our future studies.     



\section{\bfseries Our Algorithm and Its Fault-Tolerance}
\label{sec:algo}
In this section, we present a decentralized algorithm, formally stated in Algorithm \ref{algo}, for solving problem \eqref{eqn:hon_obj} under the presence of at most $f$ Byzantine faulty agents. A crucial component of our algorithm is the {\em Comparative Elimination} (CE) filter
which helps mitigate the detrimental impact of potentially incorrect information shared by the Byzantine faulty agents. The key ideas of our algorithm are further explained as follows. We let $\H \subseteq \V$, with $\mnorm{\H} \geq n-f$, denote the set of non-faulty agents.\\  


Our algorithm, Algorithm \ref{algo}, is iterative where in each iteration $t \in \{0, \, 1, \ldots \}$ each non-faulty agent $i \in \H$ maintains a variable $x^t_{i}$ as a local estimate of $x_{\H}^*$ defined in \eqref{eqn:hon_obj}. The initial local estimate $x^0_i$ is chosen to be an arbitrarily minimum point of $Q_i(x)$.
In each iteration $t\geq 0$, the non-faulty agents update their local estimates {\em synchronously} using Steps S1--S3. Recall that the network is assumed {\em complete}, i.e., there exists a bidirectional communication link between each pair of agents.\\

In Step S1, the non-faulty agents broadcast their local estimates to other agents in the network. However, Byzantine faulty agents may send different arbitrarily incorrect estimates to different agents.
In Step S2, to mitigate the detrimental impact of incorrect local estimates shared by the faulty agents, each non-faulty agent $i$ implements the CE vector filter. In particular, each non-faulty agent $i$ computes the distances between its current local estimate and the local estimates received from the other agents in $\V\setminus \{i\}$, and then sorts these distances in a non-decreasing order, with ties broken arbitrarily, as shown in~\eqref{algo:sorting}. Then, the agent $i$ eliminates $f$ (out of $n-1$) received local estimates that are $f$ farthest from its
current local estimate. We denote the set of remaining agents for each non-faulty agent $i$ in iteration $t$ by $\F_{i}^{t}$, defined in~\eqref{eqn:ce}.\\

Finally, in step S3, the non-faulty agents update their local estimates by implementing Eq.~\eqref{eqn:update}. Specifically, each non-faulty agent $i$ first computes a weighted aggregate of its current local estimate and the local estimates of agents in the set $\F_{i}^{t}$, and then projects the computed aggregate (using Euclidean projection) onto the minimum set of its local cost function $\arg \min_x Q_i(x)$. We denote, for all $i \in \H$,
\begin{align}
    \X_{i} = \arg \min_{x \in \R^d} Q_i(x) \subseteq \R^d. \label{eqn:X}
\end{align}
Note that, as $Q_i$ is convex and differentiable, $\X_{i}$ is a closed and convex set. We denote the Euclidean projection of a point $x \in \R^d$ onto $\X_i$ by $[x]_{\X_i}$. Specifically, 
\begin{align}
    [x]_{\X_i} = \arg \min_{y \in \X_i} \norm{x - y}.\label{eqn:proj}
\end{align}
As $\X_{i}$ is closed and convex, for all $x$, $[x]_{\X_i}$ is unique~\cite{boyd2004convex}. \\


\begin{algorithm}[htb!]
\SetAlgoLined
Each non-faulty agent $i \in \H$ chooses its initial estimate $x^0_i$ arbitrarily from $\X_i$. Steps in each iteration $t \in \{0, \, 1, \ldots\}$ are as follows.
\\~\\

\noindent $\diamond ~$ {\bf Step S1 (Broadcast)}: Each non-faulty agent $i$ broadcasts its current local estimate $x^t_i$ to all the other agents in $\V \setminus \{i\}$. \\~\\
    
    We let $m^t(i,j)$ denote the local estimate received by agent $i$ from agent $j \in \V \setminus \{i\}$. \\
    
    If $j \in \H$ then $m^t(i,j) = x^t_j$, else $m^t(i,j)$ may be an arbitrary vector in $\R^d$. If no value is received from a agent $j$, then $j$ must be faulty (because the system is assumed to be synchronous) -- in this case, agent $i$ assigns the ${\bf 0}$ vector to $m^t(i,j)$.
\\~\\
    
\noindent $\diamond ~$ {\bf Step S2 (CE filter}) : Each non-faulty agent $i$ filters out $f$ of its received estimates that are $f$ farthest from its current local estimate $x^t_i$. \\

Specifically, let $j_{i, k}$ denote the agent that sent local estimate, to agent $i$, which is $k$-th close to $x^t_i$, i.e., 
    \begin{align}
        \norm{x^t_i - m^t(i,j_{i,1})} \leq \ldots \leq \norm{x^t_i - m^t\left(i,j_{i,{n-1}} \right)}.\label{algo:sorting}
    \end{align}
    We define the {\em filter set} of agent $i$ to be 
    \begin{align}
        \F^t_i = \{j_{i,1}, \ldots, \, j_{i, n-f-1}\}. \label{eqn:ce}
    \end{align}
    ~
    
\noindent $\diamond ~$ {\bf Step S3 (Projected Consensus}): Each non-faulty agent $i$ updates its current local estimate $x^t_i$ to 
    \begin{align}
         x^{t+1}_i = \left[x^t_i - \eta \, \sum_{j \in \F^t_i} \left(x^t_i - m^t(i, j) \right) \right]_{\X_{i}} \label{eqn:update}
    \end{align}
    where $\eta$ is a nonnegative step size. 

\caption{Projected Consensus with CE Filter}
\label{algo}
\end{algorithm}%

{\bf Remark:} Let $v^*_i$ be the minimum value of cost function $\min_{x \in \R^d} Q_i(x)$ for all $i \in \H$. As $Q_i(y) \leq v^*_i$ if and only if $y \in \X_i$,
each agent $i$ can compute the projection $[x]_{\X_i}$, of a given point $x \in \R^d$, by solving the following convex optimization problem~\cite{boyd2004convex}: 
\begin{align}
\label{eqn:cvx_prob}
    \underset{y \in \R^d}{\text{minimum}} ~ \norm{x - y}^2, ~ \text{subject to:} ~ Q_i(y) \leq v^*_i.
\end{align}
For the special case where $Q_i(x) = \norm{A_i x - b_i}^2$ where $A_i \in \R^{n_i \times d}$ and $b_i \in \R^d$ for all $i$, the projection $[x]_{\X_i}$, of a given point $x \in \R^d$, can be obtained by solving the following quadratic programming problem:
\begin{align}
    & \underset{y \in \R^d}{\text{minimum}} ~ \norm{x - y}^2, ~ \text{subject to:} ~ A_i y - b_i = 0. \label{eqn:proj_affine}
\end{align}
From the closed-form solution of~\eqref{eqn:proj_affine}, e.g.,~\cite{meyer2000matrix}, $\forall x \in \R^d$,
\begin{align*}
    [x]_{\X^*_i} = \left(I - A^T_i \left( A_i A^T_i \right)^{-1} A_i \right) x + A^T_i \left( A_i A^T_i \right)^{-1} b_i 
\end{align*}
where $I$ is the identity matrix, and $(\cdot)^T$ denotes the transpose.\\

Next, we present our key fault-tolerance result.


%



    

\subsection{\bfseries Fault-Tolerance Property}
\label{sec:ft}
We show that our algorithm, Algorithm~\ref{algo}, obtains {\em exact fault-tolerance} (see Definition~\ref{def:eft}) under certain assumptions, provided that the necessary condition of $2f$-redundancy is satisfied. We let $\H \subseteq \V$ and $\B = \V \setminus \H$ denote the set of all non-faulty and faulty agents, respectively. Recall that $\mnorm{\H} \geq n-f$ and $\mnorm{\B} \leq f$. \\

Before we state the fault-tolerance guarantee of our algorithm, let us review below the necessity of $2f$-redundancy for {\em exact fault-tolerance}~\cite{gupta2020fault_podc, gupta2020resilience}. Recall the definition of $2f$-redundancy from Definition~\ref{def:2t_red}. \\


\def\out{\mathsf{Out}}
\def\d{\partial}
\begin{lemma}[Theorem 1 in~\cite{gupta2020fault_podc}]
\label{lem:imp}
A decentralized optimization algorithm has {\em exact fault-tolerance} \underline{only if} the set of non-faulty agents have {\em $2f$-redundancy}.
\end{lemma}
~


Our fault-tolerance result relies on the following assumptions about the non-faulty agents' cost functions~\cite{bertsekas1989parallel}. We let $Q_{\H}(x)$ denote the average non-faulty cost function, i.e., 
\begin{align}
    Q_{\H}(x) = \frac{1}{\mnorm{\H}} \sum_{i \in \H}Q_i(x), ~\quad \forall x \in \R^d.\label{eqn:def_cost_h}
\end{align}
~

\begin{assumption}[Existence]
\label{asp:finite}
We assume non-trivial existence of a solution~\eqref{eqn:hon_obj}. Specifically, there exists a point
\[x^*_{\H} \in \arg \min_{x \in \R^d} \sum_{i \in \H} Q_i(x) \text{ such that } \norm{x^*_\H} < \infty.\]
\end{assumption}
~

\begin{assumption}[Lipschitz smoothness]
\label{asp:lipschitz}
We assume that the non-faulty agents' gradients are Lispchitz continuous. Specifically, there exists a positive real value $\mu < \infty$ such that, $\forall i \in \H$,
\begin{align*}
    \norm{\nabla Q_i(x) - \nabla Q_i(y)} \leq \mu \norm{x - y}, \quad \forall x, \, y \in \R^d.
\end{align*}
\end{assumption}
~

\begin{assumption}[Strong convexity]
\label{asp:str_cvxty}
We assume that $Q_{\H}(x)$ is strongly convex. Specifically, there exists a positive real value $\lambda < \infty$ such that
\begin{align*}
    \iprod{x - y}{\nabla Q_{\H}(x) - \nabla Q_{\H}(y)} \geq \lambda \norm{x - y}^2, \quad \forall x, \, y \in \R^d.
\end{align*}
\end{assumption}
~

We state our key fault-tolerance result in Theorem~\ref{thm:main} below. Note that, under the strong convexity assumption, i.e, Assumption~\ref{asp:str_cvxty}, the aggregate non-faulty cost $\sum_{i \in \H} Q_i(x)$ has a unique minimum point:
\begin{align*}
    x^*_{\H} = \arg \min_{x \in \R^d} \sum_{i \in \H} Q_i(x). 
\end{align*}
We define a {\em fault-tolerance margin}:
\begin{align}
    \alpha =  \left(1 - \sqrt{1 + \frac{\lambda}{\mu}} \right)^2 - \frac{f}{n-f}. \label{eqn:alpha}
\end{align}



\begin{theorem}
\label{thm:main}
Suppose that Assumptions~\ref{asp:finite},~\ref{asp:lipschitz}, and~\ref{asp:str_cvxty} hold true. Consider Algorithm~\ref{algo} in Section~\ref{sec:algo}. If the non-faulty agents have $2f$-redundancy, and $\alpha > 0$, then for a small enough step-size $\eta$ in~\eqref{eqn:update} there exists $\rho \in [0, \,  1)$ such that
\begin{align*}
    \sum_{i \in \H} \norm{x^t_i - x^*_{\H}}^2 \leq \rho^t \sum_{i \in \H} \norm{x^0_i - x^*_{\H}}^2, \quad \forall t \geq 0.
\end{align*}
\end{theorem}
~

Theorem~\ref{thm:main} implies that Algorithm~\ref{algo} has {\em exact fault-tolerance} under Assumptions~\ref{asp:finite},~\ref{asp:lipschitz}, and~\ref{asp:str_cvxty}, provided that the necessary condition of $2f$-redundancy is satisfied, and the fault-tolerance margin $\alpha > 0$, i.e.,
\begin{align}
    \frac{f}{n-f} < \left(1 - \sqrt{1 + \frac{\lambda}{\mu}} \right)^2. \label{eqn:alpha_0}
\end{align}
~

\noindent Next, we present a proof of Theorem~\ref{thm:main} -- however, {\bf the reader may proceed to Section~\ref{sec:sim} without loss of continuity}.

\section{\bfseries Proof of Theorem~\ref{thm:main}}
\label{sec:proof}

In this section, we present a formal proof of Theorem~\ref{thm:main}. For convenience, we write $x^*_{\H}$ simply as $x^*$. Recall that $\H \subseteq \V$, with $\mnorm{\H} \geq n-f$, is the set of non-faulty agents. \\

Our proof relies on the following critical implications of the $2f$-redundancy property. If the non-faulty agents $\H$ have $2f$-redundancy then 
\begin{align}
    \underset{x \in \R^d}{\arg\min} \, \sum_{i \in \H} Q_i(x) \in \bigcap_{i \in \H} \underset{x \in \R^d}{\arg\min} \, Q_i(x). \label{eqn:int_min_sets}
\end{align}
Moreover, when both Assumption~\ref{asp:lipschitz} and~\ref{asp:str_cvxty} hold true, along with the $2f$-redundancy property, then 
\begin{align}
    \lambda \leq \mu. \label{eqn:lambda_mu}
\end{align}
Proofs for~\eqref{eqn:int_min_sets} and~\eqref{eqn:lambda_mu} can be found in~\cite[Appendix B]{gupta2020fault_podc}.\\

For each agent $i \in \H$, we define $V^t_i = \norm{x^t_i - x^*}^2$, and 
\begin{align}
    V^t = \sum_{i \in \H} V^t_{i}, \quad \forall t \geq 0. \label{eqn:def_vt}
\end{align}
Now, consider an arbitrary non-faulty agent $i \in \H$ and iteration $t \geq 0$. From the update law~\eqref{eqn:update}, we obtain that
\begin{align}
    V^{t+1}_i = \norm{\left[x^t_i - \eta \, \sum_{j \in \F^t_i} \left(x^t_i - m^t(i, j) \right) \right]_{\X_{i}} - x^*}^2 \label{eqn:vi_1}
\end{align}
where recall, from~\eqref{eqn:X}, that $\X_i$ denotes the minimum set $\arg \min_x Q_i(x)$. As the function $Q_i$ is convex and differentiable, $\X_i$ is a closed and convex set~\cite{boyd2004convex}. Recall from~\eqref{eqn:int_min_sets} under $2f$-redundancy, $x^* \in \X_i$ for all $i \in \H$. Thus, due the non-expansion property of Euclidean projection onto a convex set~\cite{boyd2004convex}, 
\[\norm{[x]_{\X_i} - x^*} \leq \norm{x - x^*}, \quad \forall x \in \R^d.\]
Substituting from above in~\eqref{eqn:vi_1} we obtain that
\begin{align*}
    V^{t+1}_i  \leq & \norm{x^t_i - x^* - \eta \, \sum_{j \in \F^t_i} \left(x^t_i - m^t(i, j) \right)}^2.
\end{align*}
As $\norm{x}^2 = \iprod{x}{x}$, the above implies that
\begin{align}
\begin{split}
    V^{t+1}_i  \leq & V^t_i - 2 \eta \iprod{x^t_i - x^*}{\sum_{j \in \F^t_i} \left(x^t_i - m^t(i, j) \right) } \\ \label{eqn:vi_2}
    & + \eta^2  \norm{\sum_{j \in \F^t_i} \left(x^t_i - m^t(i, j) \right)}^2.
\end{split}
\end{align}
We let
\begin{align}
    \phi^t_i = \iprod{x^t_i - x^*}{\sum_{j \in \F^t_i} \left(x^t_i - m^t(i, j) \right) }, \label{eqn:def_phi}
\end{align}
and
\begin{align}
    \psi^t_i = \norm{\sum_{j \in \F^t_i} \left(x^t_i - m^t(i, j) \right)}^2. \label{eqn:def_psi}
\end{align}
Upon substituting from~\eqref{eqn:def_phi} and~\eqref{eqn:def_psi} in~\eqref{eqn:vi_2} we obtain that
\begin{align}
    V^{t+1}_i  \leq   V^t_i - 2 \eta \, \phi^t_i + \eta^2 \, \psi^t_i. \label{eqn:vi_3}
\end{align}
Recall that $i$ is an arbitrary non-faulty agent. Thus, the above holds true for all $i \in \H$. Upon adding both sides of~\eqref{eqn:vi_3} for all $i \in \H$, and then substituting from~\eqref{eqn:def_vt}, we obtain that
\begin{align}
    V^{t+1} \leq V^t - 2 \eta \, \sum_{i \in \H} \phi^t_i + \eta^2 \, \sum_{i \in \H} \psi^t_i. \label{eqn:vi_sum}
\end{align}
In Parts I and II below, we obtain a lower bound on $\sum_{i \in \H} \phi^t_i$ and an upper bound on $\sum_{i \in \H} \psi^t_i$, respectively, in terms of $V^t$. Finally, upon substituting these bounds in~\eqref{eqn:vi_sum} we show an exponential convergence of $V^t$ to zero.\\
\noindent {\bf Part I:} For all $i \in \H$, let $\H_i = \H \setminus \{i\}$, and $\H^t_i$ be the set of non-faulty agents in the filtered set $\F^t_i$, i.e, 
\begin{align}
    \H^t_i = \F^t_i \cap \H_i. \label{eqn:def_hti}
\end{align}
As $\mnorm{\H_i} = \mnorm{\H} -1$ and $\mnorm{\F^t_i} = n-f-1$, $\forall ~ i \in \H$ and $t$,
\begin{align}
    \mnorm{\H} - f - 1 \leq \mnorm{\H^t_i} \leq n - f - 1. \label{eqn:size_hti}
\end{align}
Let $\B^t_i = \F^t_i \setminus \H^t_i$. From above we obtain that
\begin{align}
    \sum_{j \in \F^t_i} \left( x^t_i - m^t(i, j) \right) = & \sum_{j \in \H^t_i} \left( x^t_i - m^t(i, j) \right) \nonumber \\
    & + \sum_{j \in \B^t_i} \left( x^t_i - m^t(i, j) \right). \label{eqn:use_in_psi} 
\end{align}
As $\H_i = \H^t_i \cup \H_i \setminus \H^t_i$, and $m^t(i,j) = x^t_j$ for all $j \in \H_i$ from above we obtain that
\begin{align}
\begin{split} \label{eqn:sum_i}
    & \sum_{j \in \F^t_i} \left( x^t_i - m^t(i, j) \right) = \sum_{j \in \H_i} \left( x^t_i - x^t_j \right) \\
    & - \sum_{j \in \H_i \setminus \H^t_i} \left( x^t_i - x^t_j \right) + \sum_{j \in \B^t_i} \left( x^t_i - m^t(i, j) \right).
\end{split}
\end{align}
We denote
\begin{align}
    e^t_i = \sum_{j \in \B^t_i} \left( x^t_i - m^t(i, j) \right) - \sum_{j \in \H_i \setminus \H^t_i} \left( x^t_i - x^t_j \right). \label{eqn:et_i}
\end{align}
Substituting from~\eqref{eqn:et_i} in~\eqref{eqn:sum_i} we obtain that
\begin{align}
    \sum_{j \in \F^t_i} \left( x^t_i - m^t(i, j) \right) = \sum_{j \in \H_i} \left( x^t_i - x^t_j \right) + e^t_i. \label{eqn:sum_i_2}
\end{align}
Substituting from~\eqref{eqn:sum_i_2} in~\eqref{eqn:def_phi} implies that, $\forall i \in \H$,
\begin{align*}
    \phi^t_i = \iprod{x^t_i - x^*}{\sum_{j \in \H_i} \left(x^t_i - x^t_j \right) } + \iprod{x^t_i - x^*}{e^t_i}. 
\end{align*}
Therefore, 
\begin{align*}
    \sum_{i \in \H}\phi^t_i  = \sum_{i \in \H} \left(\iprod{x^t_i - x^*}{\sum_{j \in \H_i} \left(x^t_i - x^t_j \right) } + \iprod{x^t_i - x^*}{e^t_i} \right). 
\end{align*}
Note that, as $x^t_i - x^t_j = \left( x^t_i - x^* \right) - \left( x^t_j - x^* \right)$, 
\begin{align*}
    \sum_{i \in \H}\iprod{x^t_i - x^*}{\sum_{j \in \H_i} \left(x^t_i - x^t_j \right) } = \frac{1}{2}\sum_{i \in \H}\sum_{j \in \H_i}\norm{x^t_i - x^t_j}^2. 
\end{align*}
From above we obtain that
\begin{align}
    \sum_{i \in \H}\phi^t_i  = \sum_{i \in \H} S^t_i \label{eqn:sum_phi_t}
\end{align}
where
\begin{align}
    S^t_i = \frac{1}{2} \sum_{j \in \H_i} \norm{x^t_i - x^t_j}^2 + \iprod{x^t_i - x^*}{e^t_i}. \label{eqn:def_st_i}
\end{align}
From Cauchy-Schwartz inequality, we obtain that
\begin{align}
    \iprod{x^t_i - x^*}{e^t_i} \geq - \norm{x^t_i - x^*} \, \norm{e^t_i}. \label{eqn:cs_1}
\end{align}
Recall, from~\eqref{eqn:et_i}, that 
\begin{align*}
    e^t_i = \sum_{j \in \B^t_i} \left( x^t_i - m^t(i, j) \right) - \sum_{j \in \H_i \setminus \H^t_i} \left( x^t_i - x^t_j \right).
\end{align*}
Thus, from triangle inequality, $\forall i \in \H$,
\begin{align}
    \norm{e^t_i} \leq \sum_{j \in \B^t_i} \norm{x^t_i - m^t(i, j)} + \sum_{j \in \H_i \setminus \H^t_i} \norm{x^t_i - x^t_j}. \label{eqn:et_i_bnd}
\end{align}
Note that, due to the CE filter~\eqref{eqn:ce}, for each $k \in \B^t_i$ there exists a unique $j \in \H_i \setminus \H^t_i$ such that
\begin{align}
    \norm{x^t_i - m^t(i, k)} \leq \norm{x^t_i - x^t_j}. \label{eqn:byz_k_hon_j}
\end{align}
Substituting from above in~\eqref{eqn:et_i_bnd} implies that
\begin{align}
    \norm{e^t_i} \leq 2  \sum_{j \in \H_i \setminus \H^t_i} \norm{x^t_i - x^t_j}. \label{eqn:et_i_bnd_2}
\end{align}
Upon substituting from~\eqref{eqn:et_i_bnd_2} in~\eqref{eqn:cs_1} we obtain that
\begin{align}
    \iprod{x^t_i - x^*}{e^t_i} \geq - \sum_{j \in \H_i \setminus \H^t_i} 2 \norm{x^t_i - x^*} \, \norm{x^t_i - x^t_j}. \label{eqn:et_i_bnd_3}
\end{align}
As $2ab \leq 2a^2 + \frac{b^2}{2}$ for $a, \, b \in \R$,~\eqref{eqn:et_i_bnd_3} implies that
\begin{align*}
    \iprod{x^t_i - x^*}{e^t_i} \geq - \sum_{j \in \H_i \setminus \H^t_i} \left(2 \norm{x^t_i - x^*}^2 + \frac{1}{2} \norm{x^t_i - x^t_j}^2 \right). 
\end{align*}
Substituting from above in~\eqref{eqn:def_st_i} we obtain that
\begin{align}
    S^t_i \geq & \frac{1}{2} \left\{\sum_{j \in \H_i} \norm{x^t_i - x^t_j}^2 - \sum_{j \in \H_i \setminus \H^t_i} \norm{x^t_i - x^t_j}^2\right\} \nonumber \\
    & - 2 \sum_{j \in \H_i \setminus \H^t_i} \norm{x^t_i - x^*}^2 \nonumber 
\end{align}
As $\H^t_i = \H_i \setminus \{\H_i \setminus \H^t_i\}$, from above we obtain that
\begin{align}
    S^t_i \geq \frac{1}{2} \sum_{j \in \H^t_i}\norm{x^t_i - x^t_j}^2 - 2 \sum_{j \in \H_i \setminus \H^t_i} \norm{x^t_i - x^*}^2. \label{eqn:sti_2}
\end{align}
Now, we consider below the summation $\sum_{j \in \H_i^t} \norm{x^t_i - x^t_j}^2$ for an arbitrary non-faulty agent $i$ and iteration $t$. Note that, as $\norm{x^t_i - x^t_i}^2 = 0$,
\begin{align}
    \sum_{j \in \H_i^t} \norm{x^t_i - x^t_j}^2 = \sum_{j \in \H_i^t\cup\{i\}} \norm{x^t_i - x^t_j}^2. \label{eqn:xi_xj_1}
\end{align}
Lipschitz continuity of $\nabla Q_i$, i.e., Assumption~\ref{asp:lipschitz}, implies that
\begin{align*}
    \norm{x^t_i - x^t_j} \geq \frac{1}{\mu} \, \norm{\nabla Q_j(x^t_i) - \nabla Q_j(x^t_j)}, ~ \forall j \in \H^t_i.
\end{align*}
Recall, from~\eqref{eqn:update}, that for all $j \in \H$, $x^t_j \in \arg \min_x Q_j(x)$. Thus, $\nabla Q_j(x^t_j) = 0$. Substituting this above implies that
\begin{align}
    \norm{x^t_i - x^t_j} \geq \frac{1}{\mu} \, \norm{\nabla Q_j(x^t_i)}, ~ \forall j \in \H^t_i.\label{eqn:xi_xj_2}
\end{align}
Substituting from~\eqref{eqn:xi_xj_2} in~\eqref{eqn:xi_xj_1} we obtain that
\begin{align}
    \sum_{j \in \H^t_i}\norm{x^t_i - x^t_j}^2 \geq \frac{1}{\mu^2} \sum_{j \in \H^t_i \cup \{i\}} \norm{\nabla Q_j(x^t_i)}^2. \label{eqn:sum_hti_xi_xj}
\end{align}
Substituting from~\eqref{eqn:sum_hti_xi_xj} in~\eqref{eqn:sti_2} we obtain that
\begin{align*}
    S^t_i \geq \frac{1}{2\mu^2} \sum_{j \in \H^t_i \cup \{i\}} \norm{\nabla Q_j(x^t_i)}^2 - 2 \sum_{j \in \H_i \setminus \H^t_i} \norm{x^t_i - x^*}^2.
\end{align*}
As $\mnorm{\H_i \setminus \H^t_i} \leq f$ (see~\eqref{eqn:def_hti}), the above implies that
\begin{align}
    S^t_i \geq \frac{1}{2\mu^2} \sum_{j \in \H^t_i \cup \{i\}} \norm{\nabla Q_j(x^t_i)}^2 - 2 f \norm{x^t_i - x^*}^2. \label{eqn:sti_3}
\end{align}
Next, 
we obtain a lower bound on $\sum_{j \in \H^t_i \cup \{i\}} \norm{\nabla Q_j(x^t_i)}^2$ in terms of $\norm{x^t_i - x^*}^2$ for an arbitrary agent $i \in \H$ and $t$. Note that under $2f$-redundancy~\cite{gupta2020fault_podc}, $\nabla Q_j(x^*) = 0.$ for all $j \in \H$. Therefore, under $2f$-redundancy and Lipschitz continuity, i.e., Assumption~\ref{asp:lipschitz}, 
\begin{align}
    \norm{\nabla Q_j(x^t_i)} \leq \mu \norm{x^t_i - x^*}, \quad \forall j \in \H. \label{eqn:nabla_qj_xi}
\end{align}
Strong convexity of $Q_\H(x)$, i.e., Assumption~\ref{asp:str_cvxty}, implies that
\begin{align}
    \sum_{j \in \H} \iprod{x^t_i - x^*}{\nabla Q_j(x^t_i)} \geq \lambda  \mnorm{\H} \, \norm{x^t_i - x^*}^2. \label{eqn:str_1}
\end{align}
We denote, for all $i \in \H$ and $t \geq 0$,
\[\H^{t(-)}_i = \H \setminus \H^t_i \setminus \{i\}.\] 
Substituting the above in~\eqref{eqn:str_1}, we obtain that
\begin{align*}
& \sum_{j \in \H^t_i \cup \{i\}} \iprod{x^t_i - x^*}{\nabla Q_j(x^t_i)} \geq \lambda  \mnorm{\H} \, \norm{x^t_i - x^*}^2 \nonumber \\
& - \sum_{j \in \H^{t(-)}_i} \iprod{x^t_i - x^*}{\nabla Q_j(x^t_i)}. 
\end{align*}
Due to Cauchy-Schwartz inequality, 
\begin{align}
    \iprod{x^t_i - x^*}{\nabla Q_j(x^t_i)} \leq \norm{x^t_i - x^*} \, \norm{\nabla Q_j(x^t_i)}. \label{eqn:cs_ineq}
\end{align}
Thus,
\begin{align}
    & \sum_{j \in \H^t_i \cup \{i\}} \iprod{x^t_i - x^*}{\nabla Q_j(x^t_i)} \geq \lambda  \mnorm{\H} \, \norm{x^t_i - x^*}^2 \nonumber \\
    & - \mnorm{ \H^{t(-)}_i} \norm{x^t_i - x^*} \norm{\nabla Q_j(x^t_i)}. \label{eqn:nabla_qj_xi_2}
\end{align}
Substituting from~\eqref{eqn:nabla_qj_xi} in~\eqref{eqn:nabla_qj_xi_2}, we obtain that
\begin{align*}
    & \sum_{j \in \H^t_i \cup \{i\}} \iprod{x^t_i - x^*}{\nabla Q_j(x^t_i)} \\
    &\geq  \left( \lambda  \mnorm{\H} - \mu \mnorm{ \H^{t(-)}_i} \right) \, \norm{x^t_i - x^*}^2. \nonumber
\end{align*}
Substituting from~\eqref{eqn:cs_ineq} in the above, we obtain that
\begin{align*}
    \sum_{j \in \H^t_i \cup \{i\}} \norm{\nabla Q_j(x^t_i)} \geq  \left( \lambda  \mnorm{\H} - \mu \mnorm{ \H^{t(-)}_i} \right) \, \norm{x^t_i - x^*}.
\end{align*}
As $\mnorm{\H^t_i} \geq \mnorm{\H} - f - 1$ (see~\eqref{eqn:size_hti}), $\mnorm{\H^{t(-)}_i} \leq f$. Using this, and the fact that $\mnorm{\H} \geq n-f$, above implies that
\begin{align}
    \sum_{j \in \H^t_i \cup \{i\}} \norm{\nabla Q_j(x^t_i)} \geq  \left( \lambda ( n - f) - \mu f \right) \, \norm{x^t_i - x^*}. \label{eqn:pre_bnd}
\end{align}
Recall, from~\eqref{eqn:alpha}, that 
\[\alpha = \left( \sqrt{1 + \frac{\lambda}{\mu}} - 1\right)^2 -  \frac{f}{n-f}.\]
As $\alpha$ is assumed positive, 
\begin{align}
    \frac{f}{n-f} < \left( \sqrt{1 + \frac{\lambda}{\mu}} - 1\right)^2 \leq \left( 1 + \frac{\lambda}{\mu} - 1\right)^2 = \frac{\lambda^2}{\mu^2}. \label{eqn:bnd_f_n-f}
\end{align}
As $\lambda \leq \mu$ (see~\eqref{eqn:lambda_mu}),~\eqref{eqn:bnd_f_n-f} implies that $f/(n-f) < \lambda/\mu$. Thus, $\lambda ( n - f) - \mu f$ in the R.H.S.~of~\eqref{eqn:pre_bnd} is positive.
Therefore, from~\eqref{eqn:pre_bnd}, and the convexity of the square function $(\cdot)^2$, we obtain that
\begin{align*}
    \sum_{j \in \H^t_i \cup \{i\}} \norm{\nabla Q_j(x^t_i)}^2 \geq \frac{\left( \lambda ( n - f) - \mu f \right)^2}{\mnorm{\H^t_i \cup \{i\} }} \norm{x^t_i - x^*}^2. 
\end{align*}
As $\mnorm{\H^t_i} \leq n - f - 1$ (see~\eqref{eqn:size_hti}), $\mnorm{\H^t_i \cup \{i\} } \geq n-f$. Thus,
\begin{align*}
    \sum_{j \in \H^t_i \cup \{i\}} \norm{\nabla Q_j(x^t_i)}^2 \geq \frac{\left( \lambda ( n - f) - \mu f \right)^2}{n-f} \norm{x^t_i - x^*}^2.
\end{align*}
Substituting from above in~\eqref{eqn:sti_3} we obtain that
\begin{align}
     S^t_i \geq \left(\frac{\left(\lambda ( n - f) - \mu f \right)^2}{2\mu^2(n-f)}  - 2 f \right) \norm{x^t_i - x^*}^2. \label{eqn:sti_pre_beta}
\end{align}
We let
\begin{align}
    \beta = \frac{\left(\lambda ( n - f) - \mu f \right)^2}{2\mu^2(n-f)}  - 2 f. \label{eqn:beta}
\end{align}
Later, we show in~\eqref{eqn:beta_alpha} that if $\alpha > 0$ then $\beta > 0$. Substituting from~\eqref{eqn:beta} in~\eqref{eqn:sti_pre_beta} we obtain that $S^t_i \geq \beta \norm{x^t_i - x^*}^2$. Substituting this in~\eqref{eqn:sum_phi_t} implies that
\begin{align*}
    \sum_{i \in \H}\phi^t_i  \geq \beta \, \sum_{i \in \H}\norm{x^t_i - x^*}^2.
\end{align*}
As $V^t =  \sum_{i \in \H}\norm{x^t_i - x^*}^2$, from above we obtain that
\begin{align}
    \sum_{i \in \H}\phi^t_i  \geq \beta \, V^t\label{eqn:sum_phi_t_2}
\end{align}
{\bf Part II:} In this part, we obtain an upper bound on $\sum_{i \in \H} \Psi^t_i$ in terms of $V^t$ for an arbitrary iteration $t$ where recall, from~\eqref{eqn:def_psi}, that
\begin{align*}
    \psi^t_i = \norm{\sum_{j \in \F^t_i} \left(x^t_i - m^t(i, j) \right)}^2.
\end{align*}
Note that
\begin{align}
    \sum_{i \in \H}\psi^t_i \leq \left( \sum_{i \in \H} \norm{\sum_{j \in \F^t_i} \left(x^t_i - m^t(i, j) \right)} \right)^2. \label{eqn:psi_1}
\end{align}
Recall, from~\eqref{eqn:use_in_psi}, that
\begin{align*}
    \sum_{j \in \F^t_i} \left( x^t_i - m^t(i, j) \right) = & \sum_{j \in \H^t_i} \left( x^t_i - m^t(i, j) \right) \nonumber \\
    & + \sum_{j \in \B^t_i} \left( x^t_i - m^t(i, j) \right).
\end{align*}
Thus, from triangle inequality,
\begin{align}
    \norm{\sum_{j \in \F^t_i} \left( x^t_i - m^t(i, j) \right)} \leq & \sum_{j \in \H^t_i} \norm{x^t_i - m^t(i, j)} \nonumber \\
    & + \sum_{j \in \B^t_i} \norm{x^t_i - m^t(i, j)}. \label{eqn:triang_psi}
\end{align}
Recall that $m^t(i, \, j) = x^t_j$ for all $j \in \H$ and $t$. Also, recall, from~\eqref{eqn:byz_k_hon_j}, that for each $k \in \B^t_i$ there exists a unique $j \in \H_i \setminus \H^t_i$ such that $\norm{x^t_i - m^t(i, k)} \leq \norm{x^t_i - x^t_j}$. Thus, from~\eqref{eqn:triang_psi} we obtain that
\begin{align}
    \norm{\sum_{j \in \F^t_i} \left( x^t_i - m^t(i, j) \right)} \leq & \sum_{j \in \H_i} \norm{x^t_i - x^t_j}. \label{eqn:psi_all_h}
\end{align}
Using triangle inequality again, from~\eqref{eqn:psi_all_h} we obtain that
\begin{align*}
    \norm{\sum_{j \in \F^t_i} \left( x^t_i - m^t(i, j) \right)} \leq \mnorm{\H_i}\norm{x^t_i - x^*} + \sum_{j \in \H_i} \norm{x^t_j - x^*}. 
\end{align*}
As $\mnorm{\H_i} = \mnorm{\H} - 1, ~ \forall i \in \H$, from above we obtain that
\begin{align}
    & \sum_{i \in \H} \norm{\sum_{j \in \F^t_i} \left( x^t_i - m^t(i, j) \right)} \leq \left( \mnorm{\H} - 1 \right) \sum_{i \in \H} \norm{x^t_i - x^*} \nonumber \\
    & + \sum_{i \in \H} \sum_{j \in \H_i} \norm{x^t_j - x^*}.  \label{eqn:sum_psi_all_h}
\end{align}
Note that, as the network $\G$ is assumed complete,
\[\sum_{i \in \H} \sum_{j \in \H_i} \norm{x^t_j - x^*} = \left( \mnorm{\H} - 1 \right) \sum_{i \in \H} \norm{x^t_i - x^*}.\]
As $\mnorm{\H} - 1 \leq \mnorm{\H}$, substituting the above in~\eqref{eqn:sum_psi_all_h} implies that
\begin{align*}
    \sum_{i \in \H} \norm{\sum_{j \in \F^t_i} \left( x^t_i - m^t(i, j) \right)} \leq 2 \mnorm{\H} \, \sum_{i \in \H} \norm{x^t_i - x^*}. 
\end{align*}
Therefore,
\begin{align}
    & \left(\sum_{i \in \H} \norm{\sum_{j \in \F^t_i} \left( x^t_i - m^t(i, j) \right)}\right)^2 \leq 4 \mnorm{\H}^2 \left(\sum_{i \in \H} \norm{x^t_i - x^*}\right)^2 \nonumber \\
    & \leq 4 \mnorm{\H}^3 \sum_{i \in \H} \norm{x^t_i - x^*}^2. \label{eqn:sum_psi_all_h_2}
\end{align}
Recall, from~\eqref{eqn:def_vt}, that $V^t = \sum_{i \in \H} \norm{x^t_i - x^*}^2$. Substituting from~\eqref{eqn:sum_psi_all_h_2} in~\eqref{eqn:psi_1} we obtain that
\begin{align}
    \sum_{i \in \H}\psi^t_i \leq 4 \mnorm{\H}^3 \, V^t. \label{eqn:upp_bnd_psi}
\end{align}
\noindent {\bf Final step:} Recall, from~\eqref{eqn:vi_sum}, that for all $t \geq 0$,
\begin{align*}
    V^{t+1} \leq V^t - 2 \eta \, \sum_{i \in \H} \phi^t_i + \eta^2 \, \sum_{i \in \H} \psi^t_i.
\end{align*}
Substituting from~\eqref{eqn:sum_phi_t_2} and~\eqref{eqn:upp_bnd_psi} in the above we obtain that
\begin{align}
    V^{t+1} \leq \left( 1 - 2 \beta \, \eta + 4 \mnorm{\H}^3 \, \eta^2\right) V^t. \label{eqn:final_vi_sum_1}
\end{align}
Upon substituting, in~\eqref{eqn:final_vi_sum_1},
\begin{align}
    \rho = 1 - 2 \beta \, \eta + 4 \mnorm{\H}^3 \, \eta^2, \label{eqn:rho}
\end{align}
we obtain that 
\begin{align*}
    V^{t+1} \leq \rho \, V^t, \quad \forall t \geq 0.
\end{align*}
Upon retracing the above from $t$ to $0$ we obtain that
\begin{align*}
    V^{t} \leq \rho^t \, V^0, \quad \forall t \geq 0.
\end{align*}
where recall that $V^t = \sum_{i \in \H} \norm{x^t_i - x^*}^2$. We now show below that there exists $\eta \geq 0$ for which $\rho \in [0, \, 1)$. 
Let $\a = (\lambda + 2\mu)(n-f) - \mu f$, and $\b = 2 \sqrt{\mu (\mu + \lambda)} (n-f)$. Thus, by the definition of $\beta$ in~\eqref{eqn:beta}, 
\begin{align}
    \beta = (\a - \b)(\a + \b) / (2 \mu^2 (n-f)).\label{eqn:exp_beta_1}
\end{align}
From the definition of $\alpha$ in~\eqref{eqn:alpha}, we obtain that
\begin{align}
\begin{split}
    (\a - \b)/(\mu (n-f)) & = 
    \alpha, \text{ and } \label{eqn:a-b} \\
    (\a + \b)/(\mu (n-f)) &= \alpha + 4 \, \sqrt{1 + \left(\lambda/\mu\right)}.
\end{split}
\end{align}
Substituting from~\eqref{eqn:a-b} in~\eqref{eqn:exp_beta_1} we obtain that
\begin{align}
    \beta = \alpha \left(\alpha/2 + 2 \sqrt{1 + (\lambda/\mu)}\right) \left(n-f\right). \label{eqn:beta_alpha}
\end{align}
Therefore, as $n > f$ and $\alpha > 0$, $\beta > 0$. From~\eqref{eqn:rho},
\[\rho = 1 - 2 \eta \, \left( \beta - 2 \mnorm{\H}^3 \, \eta\right).\]
Thus, there exists a small enough value of $\eta$ satisfying $2 \mnorm{\H}^3 \, \eta < \beta$ for which $\rho \in [0, \, 1)$.


\section{\bfseries Experiment}
\label{sec:sim}

\begin{figure}[htb!]
\centering
\centering \includegraphics[width=0.35\textwidth]{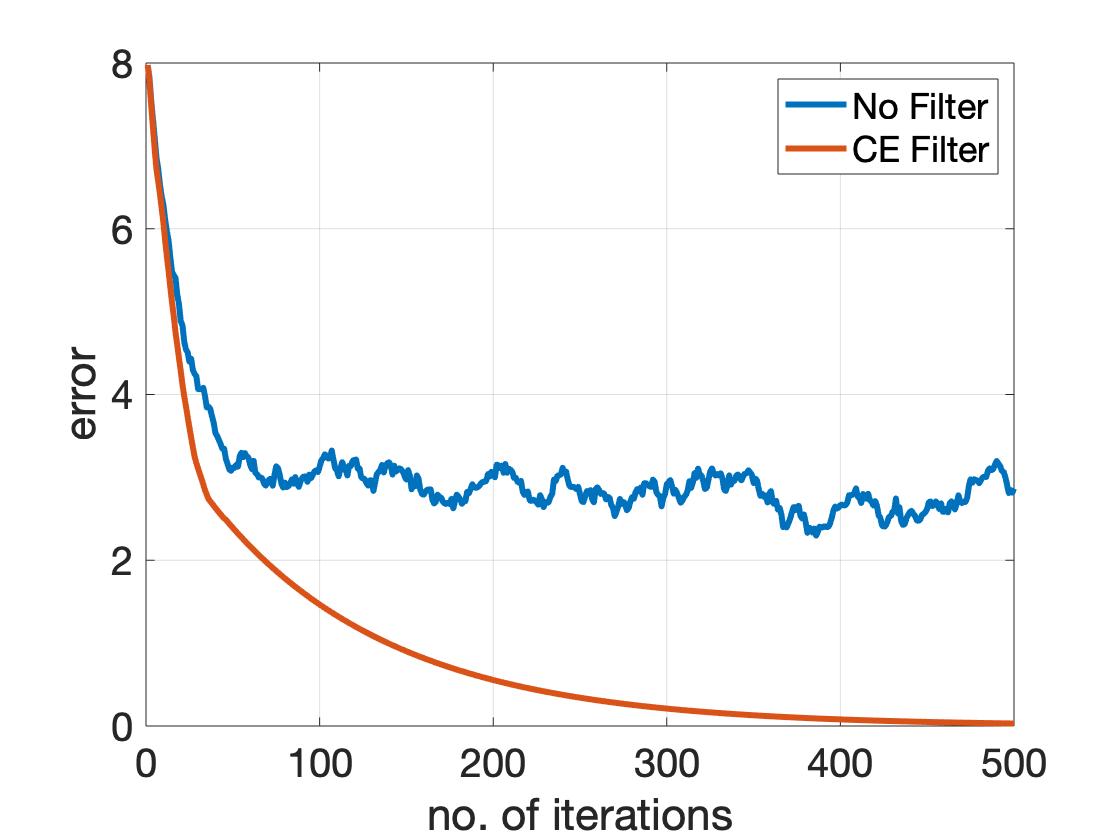} 
\caption{\footnotesize{\it Error $\sum_{i \in \H}\norm{x^t_i - x^*_\H}^2$ versus the number of iterations $t$.}}
\label{fig:sim}
\end{figure}

In this section, we present an empirical fault-tolerance result for Algorithm~\ref{algo}. We consider a complete peer-to-peer network of $n = 6$ agents with $f = 1$ Byzantine faulty agent. The cost function of each agent $i \in \{1, \ldots, \, 6\}$ is defined to be $Q_i(x) = \norm{A_i x - b_i}^2$ where $x \in \R^2$, and $A_i \in \R^{2 \times 1}$ and $b_i \in \R$ are the respective rows and elements of matrix $A$ and vector $b$ defined below.
\begin{align*}
    A &= \left[ \begin{array}{cccccc}1 & 0.8 & 0.5 & 0.3 & 1 & 0 \\ 0 & 0.5 & 0.8 & 1 & 0.3 & 1 \end{array}\right]^T, \text{ and}\\
    b &= \left[\begin{array}{cccccc} 1 & 1.3 & 1.3 & 1.3 & 1.3 & 1 \end{array}\right]^T
\end{align*}
where $(\cdot)^T$ denotes the transpose.
For our experiment, we assume that agent $6$ is Byzantine faulty, and thus, $\H = \{1, \ldots, \, 5\}$. Note that the minimum point of the aggregate cost of non-faulty agents is $x^*_\H = [1 ~ 1]^T$, and Assumptions~\ref{asp:finite},~\ref{asp:lipschitz} and~\ref{asp:str_cvxty} hold true. The non-faulty agents execute Algorithm~\ref{algo} with $\eta = 0.01$ in~\eqref{eqn:update}. In each iteration, the Byzantine agent sends different random $2$-dimensional vectors, whose elements are chosen independently and uniformly from $[0, ~ 10]$, to different agents. We observe that Algorithm~\ref{algo}, i.e., projected consensus method with CE filter, outputs the true solution $x^*_{\H}$, unlike the traditional projected consensus method without any filter~\cite{nedic2010constrained}, as shown in Fig.~\ref{fig:sim}. 

\section{\bfseries Summary}
\label{sec:sum}
In this paper, we have proposed a Byzantine fault-tolerant multi-agent decentralized optimization algorithm. We have shown that our algorithm obtains {\em exact fault-tolerance} against up to $f$ Byzantine faulty agents in a complete peer-to-peer network, provided the non-faulty agents satisfy the necessary condition of $2f$-redundancy, and the fraction of faulty to non-faulty agents $f/(n-f)$ is bounded.

\bibliographystyle{IEEEtran}
\bibliography{ref.bib}

\end{document}